\documentclass[apj]{emulateapj}
\usepackage{graphics,graphicx}
\usepackage{epsfig}
\usepackage{dcolumn}
\usepackage{rotating}
\usepackage{lscape}
\shorttitle{A new X-ray method to measure the mass of black holes}
\shortauthors{Gliozzi et al.}

\def\mbh{$M_{\rm BH}$}

  \def\pks{PKS~0558--504}

  \def\chandra{{\it Chandra}} 
  \def\xmm{{\it XMM-Newton}}

  \def\swift{{\it SWIFT}}

  \def\lum{erg s$^{-1}$}

  \def\arcsec{$^{\prime\prime}$}

  \def\ltsima{$\; \buildrel < \over \sim \;$}
  \def\simlt{\lower.5ex\hbox{\ltsima}} % < over ~
  \def\gtsima{$\; \buildrel > \over \sim \;$}
  \def\simgt{\lower.5ex\hbox{\gtsima}} % > over ~

\begin{document}
\title{Testing a scale-independent method to measure the mass of black holes}

\author{M. Gliozzi}
\affil{George Mason University, 4400 University Drive, Fairfax, VA 22030}

\author{L. Titarchuk}
\affil{George Mason University, 4400 University Drive, Fairfax, VA 22030}

\author{S. Satyapal}
\affil{George Mason University, 4400 University Drive, Fairfax, VA 22030}

\author{D. Price}
\affil{George Mason University, 4400 University Drive, Fairfax, VA 22030}

\author{I. Jang}
\affil{George Mason University, 4400 University Drive, Fairfax, VA 22030}

\begin{abstract}
Estimating the black hole mass at the center of galaxies is a fundamental step
not only for understanding the physics of accretion, but also for the cosmological
evolution of galaxies. Recently a new method, based solely on X-ray data,
was successfully applied to determine the black hole mass in Galactic systems. Since
X-rays are thought to be produced via Comptonization process both in stellar
and supermassive black holes, in principle, the same method may be applied to estimate
the mass in supermassive black holes. In this work we test this hypothesis by
 performing a systematic analysis of a sample of AGNs, whose black hole mass has been 
already determined via reverberation mapping and which possess high quality XMM-Newton archival 
data. The good agreement obtained between the black hole masses derived with
this novel scaling technique and the reverberation mapping values suggests that this
method is robust and works equally well on stellar and supermassive  black holes,
making it a truly scale-independent technique for black hole determination. 
\end{abstract}


%%%%%%%%%%%%%%%%%%%%%%%%%%%%%%%%%%%%%%%%%%%%%%%%%%%%%%%%%%%%%%%%%%%%%%%%%%%%%%%%%
%%%%%%%%%%%%%%%%%%%%%%%%%%%% SECTION 1 INTRODUCTION           %%%%%%%%%%%%%%%%%%%
%%%%%%%%%%%%%%%%%%%%%%%%%%%%%%%%%%%%%%%%%%%%%%%%%%%%%%%%%%%%%%%%%%%%%%%%%%%%%%%%%
\section{Introduction}
There is now convincing evidence that the 
most powerful persistent sources in our Galaxy (Galactic black hole systems;
hereafter GBHs) and in the universe at large (active galactic nuclei; AGNs) 
are powered by gravitational accretion onto stellar and
supermassive black holes, respectively. One of the most evident 
manifestations of black hole systems (BHs) is the presence of relativistic jets,
which may have a huge impact on their environment over distances that are
well beyond the radius of influence of the black hole itself, as 
demonstrated by recent work on the effects of powerful jets at the center of
galaxy clusters \citep[e.g.][]{mcna00,fab03}.
In addition, and perhaps more importantly,
the tight correlations found between the BH mass (\mbh) and several galaxy 
parameters such as the velocity dispersion or the mass of the bulge 
\citep[e.g.][]{gebh00,ferr00}
clearly demonstrate that black hole growth and the build-up process of 
galaxies are closely related and therefore black holes are essential 
ingredients in the evolution of galaxies. 

The estimate of the mass of supermassive black holes at the center
of galaxies is the most crucial parameter needed to understand 
the formation and evolution
processes in galaxies as well as the central engine in AGNs.
Different methods have been developed to measure \mbh\ in AGNs, 
depending on
their characteristics (e.g., degree of activity, variability, distance). 
For example, in the case of weakly active or quiescent galaxies, \mbh\ can be
determined by directly modeling the dynamics of gas or stars in the 
vicinity of the 
black hole \citep[e.g.][]{korm95,mago98}. 
On the other hand, for highly active broad-lined galaxies that show strong
variability, the most widely used way to estimate \mbh\ relies upon a technique
known as reverberation mapping (e.g., \citealt{peter04} and references therein;
see \citealt{czer10} for a recent review of the
most commonly used methods).

Mass estimates obtained by these methods have provided very important 
results for our understanding of BH formation and galaxy evolution. 
However, all methods have specific limitations. For example, methods based 
on the modeling of gas and stellar dynamics require the sphere of influence 
of the BH to be resolved. Therefore, these techniques can successfully be 
applied only to nearby galaxies with relatively massive BHs and where the galaxy's
optical emission is not substantially affected by the 
black hole activity. On the other hand, the reverberation method 
is time intensive, and cannot be applied to very luminous
sources, whose variability is typically characterized by small-amplitude
flux changes occurring on very long timescales. Furthermore, this technique 
requires the presence of a detected broad line region (BLR), whose nature and 
geometry is still poorly known. Similar limitations affect most of the secondary 
methods, which rely on some empirical relationship between \mbh\ and different 
properties of the host galaxy  (see \citealt{vest09} for a recent review).
Additionally,
bright type 1 AGNs (i.e. the AGNs characterized by the presence of a BLR) comprise 
only a small minority
of the total galaxy population, severely limiting the number of galaxies 
that can be probed with this technique.  Indeed, even amongst the population 
of  known optically-identified AGNs, type 2 AGNs (i.e., the AGNs without visible
BLR) appear to be about 4 times more numerous than type 1 AGN \citep{maio95}.

Exploring the SMBH mass function in as diverse as possible 
galactic environments is essential to understand the evolutionary history of 
BHs and their connection to their host galaxies.  To address this serious 
deficiency, it is critical to explore alternative methods to determine the BH 
mass that are not dependent on measurements at optical wavelengths.

X-ray observations are one of the most effective means to investigate the 
properties of AGNs for several reasons. 
First, the X-ray
emission is one of the defining properties of AGNs: while optical lines
generally attributed to AGN activity may be obscured, suppressed, or 
undetected for several different reasons (e.g., low signal-to-noise ratio S/N,
excessive redshift, galaxy or star light contamination), the X-ray emission
appears to be ubiquitous in AGNs. Second, the X-rays are 
produced and reprocessed in the inner, hottest nuclear regions of the source.
Therefore, unlike the optical lines from the broad and narrow line regions 
that are produced 
by the reprocessing of the primary emission, the X-ray emission directly 
traces the black hole activity. Finally, the penetrating power of (hard) X-rays 
allows them to carry information from the inner core regions without being 
substantially affected by absorption. 

At first order, from the spectral point of view, the AGN X-ray emission in the 
2--10 keV energy range is adequately described by a simple power-law model.
Importantly, the photon index $\Gamma$ appears to correlate directly with
the accretion rate, suggesting that X-ray spectral properties provide
indirect information on the accretion state in AGNs.
While there is growing evidence of a positive correlation between
$\Gamma$ and Eddington ratio for bright AGNs \citep[e.g.][]{shem06,papa09},
there is also suggestive evidence of a possible anti-correlation for BH systems
accreting at low Eddington ratio \citep[e.g.][]{const09, wugu08}.
Although several details are still unknown,
it is now widely accepted that the bulk of the X-ray emission both in AGNs and
GBHs is produced by the Comptonization process, i.e. the repeated inverse 
Compton scattering, where seed photons 
likely produced via thermal emission by an accretion disk
are up-scattered by highly energetic electrons. 
Therefore, X-rays represent one of the most effective tools to probe
the properties of black holes systems and may provide independent
constraints on their mass.
 
In this work we investigate whether a novel method to estimate \mbh, which 
is based solely 
on X-ray spectral data and was successfully applied to GBHs by 
Shaposhnikov \& Titarchuk (2009; hereafter ST09), can be
extended to AGNs. To this end, we utilize a sample of AGNs with \mbh\ already
well constrained via reverberation mapping and with good-quality X-ray data.
In Section 2 we summarize the main characteristics of this method. The
AGN sample and the data reduction are described in Section 3.
The results are presented in Section 4 and discussed in Section 5, where we
draw our main conclusions.

%%%%%%%%%%%%%%%%%%%%%%%%%%%%%%%%%%%%%%%%%%%%%%%%%%%%%%%%%%%%%%%%%%%%%%%%%%%%%%%%%
%%%%%%%%%%%%%%%%%%%%%%%%%%%% A NEW METHOD TO ESTIMATE MBH     %%%%%%%%%%%%%%%%%%%
%%%%%%%%%%%%%%%%%%%%%%%%%%%%%%%%%%%%%%%%%%%%%%%%%%%%%%%%%%%%%%%%%%%%%%%%%%%%%%%%%
\section{A new method to estimate \mbh\ }
Recently \citet{shap09} carried out an extensive and systematic analysis
of the temporal and spectral X-ray data from 17 spectral transition episodes 
in 8 different GBHs. Their main findings can be summarized as follows: 
(1) Two positive correlations are found: the first one involving temporal and
spectral properties and specifically relating the quasi periodic 
oscillation (QPO) frequency and the photon index $\Gamma$; the second one 
relating two spectral parameters of a specific Comptonization model described below: 
the normalization $N_{\rm BMC}$ and $\Gamma$.
(2) Both 
spectral evolution trends (QPO - $\Gamma$ and $N_{\rm BMC}$ - $\Gamma$) can
be adequately parametrized by analytical functions, which are similar
for the different GBHs. (3) The self-similarity of these pair of trends  
in different GBHs allows the estimate of \mbh\ and distance for these binary systems.

Since the Comptonization process producing the X-ray emission is widely 
thought to work in the same way in stellar and supermassive black holes, 
in principle, the same method may be applied to estimate the mass of 
supermassive black holes. Clearly, it is not possible to use the 
QPO - $\Gamma$ diagram, since no firm detections of QPOs  have been 
reported in AGNs (with the exception of the NLS1 RE~J1034+396
in one occasion; \citealt{gier08}). But this does not represent
a problem for AGNs, since their distance is independently and robustly determined 
from redshift measurements or from variable star methods for objects of 
the Local Group.
Hereafter, we will therefore focus solely on the $N_{\rm BMC}$ - $\Gamma$ 
diagram. 

In the following, we first describe the general details of the Comptonization model
used to fit the X-ray spectra and the basic characteristics of the scaling method that 
allows one to estimate the mass for any BH system by simply scaling the
\mbh\  value from a suitable Galactic reference source. 

%%%%%%%%%%%%%%%%%%%%%%%%%%%% SUBSECTION 2.1 BMC MODEL    %%%%%%%%%%%%%%%%%%%
\subsection{The BMC model}
Although the geometry and nature of the Comptonizing region
are still matter of debate because of spectral degeneracy, 
for our purpose (i.e. estimating black hole masses), 
the specific physical condition of the Comptonizing medium (i.e. energy and 
spatial distribution of the upscattering electrons and their optical depth) are
negligible, provided that the Comptonization process is 
characterized in a simple, general, 
yet physically correct way. These criteria are met by the so called
Bulk Motion Comptonization model (BMC), which  is a generic Comptonization 
model able to describe equally well the thermal Comptonization (i.e. the 
inverse Compton scattering produced by electrons with a Maxwellian energy 
distribution)
and the bulk motion Comptonization (where the seed photons are scattered off
electrons with bulk relativistic motion),
although it was historically developed to describe the Comptonization of 
thermal seed photons by a relativistic converging flow \citep{tita97}.

The BMC model is characterized by 4 free parameters: 1) the temperature of the
thermal seed photons $kT$, 2) the energy spectral index $\alpha$ (which is
related to the photon index by the relation $\Gamma=1+\alpha$), 3) a parameter 
$\log(A)$ related to the Comptonization fraction $f$ (.i.e., the ratio
between the number of Compton scattered photons and the number of seed 
photons) by the relation
$f=A/(1+A)$, and 4) the normalization $N_{\rm BMC}$.

In simple words, the BMC model (which is implemented in the spectral fitting
package \verb+Xspec+)
convolves the thermal seed photons 
and a generic Comptonization Green's function producing
a power law. As a consequence, this spectral model generally provides 
a good fit for X-ray spectra of accreting black holes, since 
their continuum is adequately described by a power law.

The BMC model presents two important advantages with respect to the power-law
model (PL), which is often used to parametrize the Comptonization
component: 1) Unlike the PL, which is a phenomenological model, the BMC 
parameters
are computed in a self-consistent way; 2) Unlike the PL, the power law produced
by BMC does not extend to arbitrarily low energies and thus does not affect
the normalization of the thermal component nor the amount of local 
absorption, which is often present around accreting objects.

%%%%%%%%%%%%%%%%%%%%%%%%%%%% SUBSECTION 2.2 A NEW SCALING METHOD %%%%%%%%%%%%%%%%%%%
\subsection{A new scaling method}
Generally, during their transitions from the low-hard (LH) to the 
high-soft (HS) spectral state different GBHs show similar spectral variability 
patterns. 
Specifically, when the X-ray spectra are fitted with the BMC model, and 
$\Gamma$ is plotted versus $N_{\rm BMC}$, the spectral evolution appears to
be characterized by two saturation levels (the lower one corresponding to the 
LS and the higher one to the HS) connected by a monotonically increasing curve. 
Following ST09, this spectral trend can be parameterized by the following 
functional form:
\begin{equation}
\Gamma(N_{\rm BMC}) = A - B \ln\left\{\exp\left[1-(N_{\rm BMC}/N_{\rm tr})^\beta\right]+1\right\}
\end{equation}
Note that ST09 provide in their Equation (10) a slightly different 
functional form that comprises
an additional parameter $D$ in order to be able to fit both the QPO - $\Gamma$
and $N_{\rm BMC}$ - $\Gamma$ trends. However, since $D=1$ for the pattern of 
our interest, Equation (1) coincides with Eq.(10) of ST09.

By virtue of the similarity of these $N_{\rm BMC}$ - $\Gamma$ trends among GBHs,
the value of \mbh\ (and the distance, when the $N_{\rm BMC}$ - $\Gamma$
plot is used in combination with the 
QPO - $\Gamma$ plot) of any GBH can be obtained from 
a direct scaling process. Simply speaking, if \mbh\ (and the distance) is known 
for a suitable GBH considered as a reference system, the black hole mass 
for any 
other GBH can be determined by horizontally shifting its self-similar function 
until it matches the reference object's pattern (a visual explanation of this
scaling process is provided in Section 3 when this technique is applied to AGNs). 
%%--------------FIG1-------functional form-------------------------
\begin{figure*}
\begin{center}
\includegraphics[bb=100 85 550 760,clip=,angle=0,width=14cm]{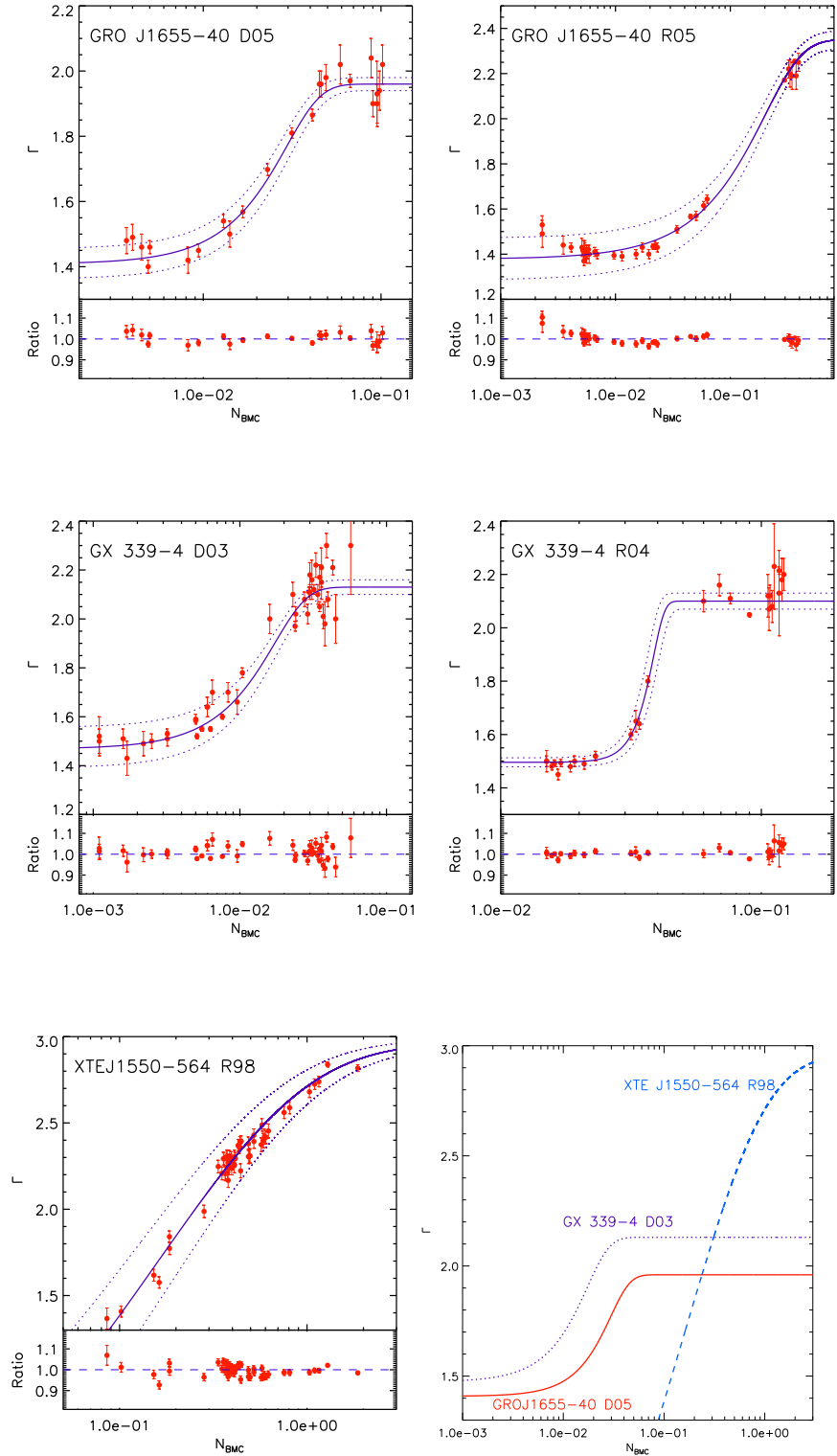}
\end{center}
\caption{\footnotesize
Photon index measurements plotted versus the BMC normalization values
for different transitions of our three reference sources GRO J1655-40, GX 339-4, and 
XTE J1550-564. The thick solid line
represents the best fit, whereas the dotted lines are the 1$\sigma$ confidence
levels. The small bottom panels describe the data-to-model
ratios. Details on the fitting procedure and the values of the functional parameters
are provided in the text and in Table 2. The right bottom panel shows a
comparison between the spectral variability trends of the three reference sources
on the same scale.
}
\label{fig1}
\end{figure*}
%%-----------------------------------------------------------------

The physical basis of these scaling techniques is thoroughly explained by
ST09 and essentially relies on few
reasonable considerations that can be summarized as
follows: (1) the QPO frequency is inversely proportional to \mbh\ (which can
be readily understood considering that
the larger \mbh, the larger the distance from the BH of the accretion
inflow responsible for the radiation, and hence the longer the dynamical
timescale of the process); (2) the BMC normalization is a function of 
luminosity and distance: $N_{\rm BMC}\propto L/d^2$ (by definition of the BMC
model); 
(3) the luminosity of an accreting black hole system can be expressed as 
$L\propto \eta M_{\rm BH} \dot m$, where $\eta$ is the radiative efficiency
and $\dot m$ the accretion rate in Eddington units.
The self-similarity of the $N_{\rm BMC}$ - $\Gamma$  correlation 
implies that, in the same spectral state, different sources have similar values of 
$\eta$ and $\dot m$. Therefore, the photon index is a reliable indicator for the 
source's spectral state independently of its BH mass.

In simple terms, the necessary steps to derive \mbh\ with this method
can be summarized as follows:\\
\noindent (1) Construct a $\Gamma -
N_{\rm BMC}$ plot for a GBH of known mass and distance, which will be
used as reference (hereafter denoted by the subscript {\it r})
and for a target of interest (denoted by the subscript {\it t}).\\
\noindent (2) Infer the normalization ratio between the target and the 
reference object $N_{\rm BMC,t}/N_{\rm BMC,r}$ by shifting in the $\Gamma -
N_{\rm BMC}$ plot the target's
pattern until it matches the reference one.\\
%at the value of $\Gamma$ measured for the target.
\noindent (3) Derive the black hole mass using the following equation
\begin{equation}
M_{\rm BH,t}=M_{\rm BH,r}\times 
(N_{\rm BMC,t}/N_{\rm BMC,r})\times (d_t/d_r)^2 \times f_G
\end{equation}
where $M_{\rm BH,r}$ is the black hole mass of the reference object,
$N_{\rm BMC,t}$ and $N_{\rm BMC,r}$ are the respective BMC normalizations
for target and reference objects, $d_t$ and $d_r$ are the
corresponding distances, and $f_G=\cos\theta_r/\cos\theta_t$ is a
geometrical factor that depends on the respective inclination angles
and should be included only in the scenario where the X-ray 
soft photon emitting region has a disk-like geometry.

Following ST09, we assume GRO J1655-40 as our primary reference source,
since its system's parameters are the most tightly constrained.
However, for completeness, we will consider all the reference sources
and patterns used by ST09.
The basic properties of these GBHs are reported in
Table 1.

In Figure~\ref{fig1} we show the $\Gamma-N_{\rm BMC}$ plots for our three 
reference sources during rise and decay phases of different outbursts 
(the details of these transitions are provided in ST09 and \citealt{tita10}).
Although similar at first sight, the spectral patterns differ from 
source to source as evidenced by the bottom right panel of Fig.~\ref{fig1}.
However, one must bear in mind that the $\Gamma-N_{\rm BMC}$ plot
yields the $N_{\rm BMC,t}/N_{\rm BMC,r}$ ratio, but the
\mbh\ estimate also depends on the values of mass and distance
of the reference source (see Eq. 2).
As a consequence, reference sources with different patterns 
(and different mass and distance) may lead to
similar estimates of \mbh\ (see Section 4.2).

From Figure~\ref{fig1} we also note that the same source displays different 
spectral patterns during the rise  and the the decay phase of an outburst. 
The latter behavior simply reflects the observational fact that during 
any outburst every GBH shows an hysteresis loop when the spectral properties
(in this case the photon index $\Gamma$) are plotted versus the its intensity
(here parameterized by the normalization $N_{\rm BMC}$). 

As outlined by 
ST09, when this method is utilized to estimate \mbh\ in a given GBH, one 
must use the appropriate reference plot. For example, a $\Gamma-N_{\rm BMC}$ plot
obtained during an outburst rising phase for a given
target source can only be compared to a rising pattern of a suitable reference
source that shows a similar trend. When applying this method
to AGNs, given the  much longer timescales involved, we cannot construct a complete
pattern in the $\Gamma-N_{\rm BMC}$ plot nor determine whether the AGN is in
a rising or decaying outburst phase. Therefore, for completeness, we need to use
all the available reference patterns and then determine which one provides the 
best match with the values obtained from reverberation mapping.

We fitted the X-ray spectral trends with the
functional form described in Equation (1) using the Levenberg-Marquardt 
algorithm \citep{press97}. The results of the fit, which are reported in 
Table 2 along with the 1$\sigma$ uncertainties on each parameter,
are broadly consistent with those of ST09, the main difference 
being that in our fits all parameters are left free to vary. In this way, 
we can determine the uncertainty on each parameter, which is then taken into 
account to estimate the uncertainty on the \mbh\ values.

A priori, any of the patterns shown in Figure~\ref{fig1} can equally
well be used to determine \mbh\ in our sample of AGNs. Indeed, all three 
reference sources have specific advantages: GRO J1655-40 has the best determined
system parameters and therefore reduces the uncertainty associated with the 
estimate of \mbh; GX 339-4 is known to be a prototypical GBH from the
spectral variability point of view with consistent patterns in all the
outbursts observed; finally, XTE J1550-564 displays the largest range of 
photon indices during its spectral transition, allowing to extend the 
estimate of \mbh\ to AGNs with steep $\Gamma$. However,
a visual inspection of Figure~\ref{fig1} reveals that 
the patterns describing the outburst rise phase for the three reference
sources (top and middle right panels and bottom left panel)
lack the coverage of at least one part of the spectral
pattern: part of the rising trend and all the upper saturation level for 
GRO J1655, the rising trend for GX 339-4, and both saturation levels 
for XTEJ1550-564. On the other hand, the decay phases of GRO J1655-40 and 
GX 339-4 (hereafter GRO J1655 D05 and GX 339 D03, respectively) are well-sampled across the entire range of the spectral pattern, and hence may be considered
a priori more reliable. Nevertheless, only a posteriori, after
a comparison of the \mbh\ estimates with the corresponding values obtained
from reverberation mapping will we be able to assess which reference pattern
is preferable and in which context.

%%%%%%%%%%%%%%%%%%%%%%%%%%%%%%%%%%%%%%%%%%%%%%%%%%%%%%%%%%%%%%%%%%%%%%%%%%%%%%%%%
%%%%%%%%%%%%%%%%%%%%%%%%%%%% SECTION 3    TEST ON AGN         %%%%%%%%%%%%%%%%%%%
%%%%%%%%%%%%%%%%%%%%%%%%%%%%%%%%%%%%%%%%%%%%%%%%%%%%%%%%%%%%%%%%%%%%%%%%%%%%%%%%%

\section{Test of the scaling method on active galactic nuclei}
Despite the large difference in scales, both GBHs and AGNs are believed to
harbor the same central engine: a black hole and an accretion disk/corona
that sometimes produces relativistic jets.
There is now mounting evidence that AGNs may be considered as
large-scale analogs of GBHs \citep[see, e.g.,][]{koerd06,mchar06,sobol09}. 
Therefore, the progress made 
in the field of GBHs can be in principle extended to AGNs (and vice versa).

Because of their higher brightness (due to their vicinity) and  their
shorter variability timescales (direct consequence of their smaller \mbh), 
the temporal and spectral properties of GBHs are much better known and can 
be used to infer information on their more powerful, extragalactic analogs.
In the framework of the AGN-GBH unification, it thus appears
reasonable to extend to AGNs the scaling method described before to 
determine \mbh. 

\begin{figure}[ht]
\centering
\includegraphics[bb=40 30 450 350,clip=,angle=0,height=6.cm]{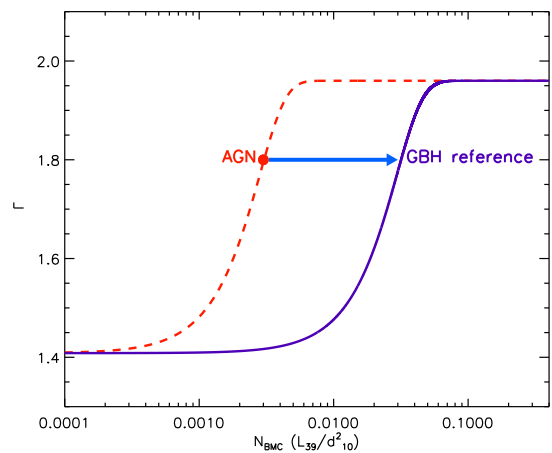}
\caption{\footnotesize  $N_{\rm BMC}$ - $\Gamma$
diagram for a GBH reference source and for an hypothetical AGN. The thick
continuous line represents the best fitting function of the GBH spectral
trend.  The arrow illustrates how we determine the value of
$N_{\rm BMC,t}/N_{\rm BMC,r}$  from the value of $\Gamma$ measured for the AGN.
See text for further details. 
}
\label{fig2}
\end{figure}

To illustrate how this method can be extended to AGNs, in Figure~\ref{fig2} 
we show the $\Gamma - N_{\rm BMC}$ diagram for a GBH reference source  
(thick solid line), and for an hypothetical AGN with $\Gamma=1.8$ 
(filled circle). The only basic assumption made is that, on 
a much longer timescale than GBHs,
the AGN follows a similar spectral pattern (light dashed line).
The scaling factor necessary to obtain $M_{\rm BH,AGN}$
starting from a reference value $M_{\rm BH,GBH}$ is given by the product
$(N_{\rm BMC,t}/N_{\rm BMC,r})\times (d_{\rm t}/d_{\rm r})^2$, 
where the first factor is determined by 
shifting right-ward the dashed gray (red, if printed in color) trend 
until it matches the solid darker one. In other words the difference 
in \mbh\ between the target of interest and the reference system 
is directly related to the amount of the shift along the x-axis
in the $\Gamma - N_{\rm BMC}$ diagram and to the square of the
distance ratio.
%%%----------FIG2 N_BMC-Gamma plot---------------------
%%%----------------------------------------

%%%%%%%%%%%%%%%%%%%%%%%%%%%% SUBSECTION 3.1 AGN SAMPLE   %%%%%%%%%%%%%%%%%%%
\subsection{The AGN sample}
We chose a sample of AGNs
that fulfills two basic criteria: (1) the AGNs
have BH mass estimates derived from a primary method, and (2) possess
high quality X-ray data, in order to tightly constrain the parameters of 
the BMC model. Nearly 30 objects with \mbh\ determined via 
reverberation mapping by \citet{peter04} have been observed by \xmm, 
which, thanks to its high throughput in the 0.3--10 keV energy band, 
represents the best X-ray satellite for this kind of analysis.

The main properties of this sample, reported in Table 3,  can be summarized 
as follows: 
the sample spans a redshift range of 0.002--0.234; the values of \mbh\  
encompass nearly 3
orders of magnitude with log(\mbh) ranging between 6.23 and 9.11 solar masses;
the bolometric luminosities, computed by integrating the spectral
energy distribution over the 0.001--100 keV interval by \citet{vasu09}, 
span nearly 5 orders of magnitudes ranging between 
$4\times10^{42} - 1.3\times10^{47}$ \lum; finally,
the Eddington ratio $\lambda_{\rm Edd}$, obtained from the ratio between 
bolometric and Eddington luminosity, covers the range between 0.001--1.14.
In summary, not only this AGN sample fulfills the 2 basic criteria, but also
all the relevant physical parameters (with particular emphasis for
\mbh\ and the Eddington ratio, which 
are crucial for the comparison between AGN and GBHs) span a considerably 
large region of the parameter space, providing the ideal framework
to test the scaling method.

%%%%%%%%%%%%%%%%%%%%%%%%%%%% SUBSECTION 3.2 XMM DATA REDUCTION   %%%%%%%%%%%%%%%%%%%
\subsection{Data reduction}
Since basically all sources are bright and with relatively large exposures
(the \xmm\ observation identifiers and the net exposures are reported in
Table 4), we restrict 
our analysis to the EPIC pn data, which provide the highest S/N in the energy
band of interest. The EPIC  data were processed in a homogeneous way
using the latest CCD gain
values. For the temporal and spectral analysis, events corresponding to pattern 
0--4 (singles and doubles only) were accepted. 
Arf and rmf files were created with the \xmm\ Science 
Analysis Software (\verb+SAS+) 8.0.
The recorded events were screened to remove known hot pixels and data affected
by periods of flaring background.  In general, the extraction radius used 
for the source spectra and light curves is 30\arcsec, whereas
background spectra and light curves were extracted from
source-free circular regions of 60\arcsec\ extraction radius
on the same chip as the source.  

The X-ray spectral analysis  was performed using the {\tt XSPEC v.12.4.0}
software package \citep{arn96}.  The EPIC data have been re-binned in order 
to contain at least 20 counts per channel for the $\chi^2$ statistic to be valid. 
The errors on spectral parameters are at 90\% confidence level for one 
interesting parameter ($\Delta \chi^2=2.71$).

%%%----------FIG3 Histogram of Eddington ratio ---------------------
\begin{figure}[ht]
\centering
\includegraphics[bb=5 5 255 235,clip=,angle=0,width=9.cm]{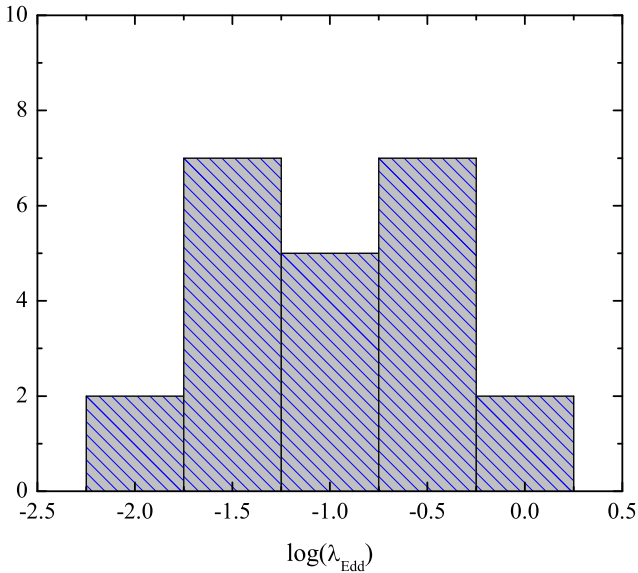}
\caption{\footnotesize  Distribution of the Eddington ratio values
$\lambda_{\rm Edd}= L_{\rm bol}/L_{\rm Edd}$ of the
AGNs for which the \mbh\ was determined with the X-ray scaling method.
}
\label{fig3}
\end{figure} 
%%%----------------------------------------

%%%%%%%%%%%%%%%%%%%%%%%%%%%%%%%%%%%%%%%%%%%%%%%%%%%%%%%%%%%%%%%%%%%%%%%%%%%%%%%%%
%%%%%%%%%%%%%%%%%%%%%%%%%%%% SECTION 4 RESULTS & DISCUSSION   %%%%%%%%%%%%%%%%%%%
%%%%%%%%%%%%%%%%%%%%%%%%%%%%%%%%%%%%%%%%%%%%%%%%%%%%%%%%%%%%%%%%%%%%%%%%%%%%%%%%%
\section{Results}
\subsection{Spectral Analysis}

We carried out a systematic spectral analysis of the X-ray data of all AGNs
listed in Table 3 using a baseline model that comprises a BMC model and two
absorption components, one fixed at the Galactic value and one free to vary to
mimic the intrinsic local absorption. When necessary, one or two Gaussian 
components were added to account for line-like features. 

%%%----------FIG4 Histogram of BH masses ---------------------
\begin{figure}[ht]
\centering
\includegraphics[bb=5 10 205 120,clip=,angle=0,width=9.cm]{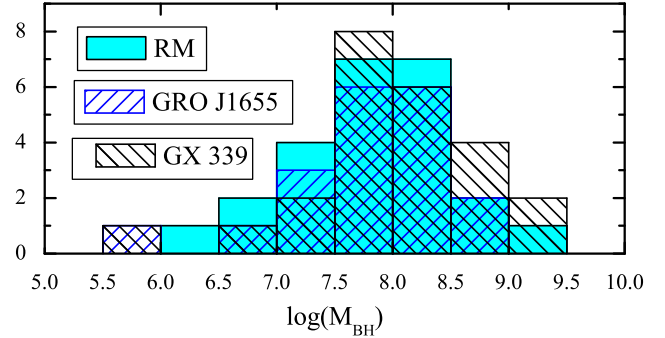}
\caption{\footnotesize   Distributions of the \mbh\ values obtained
with reverberation mapping (filled histogram), and with the X-ray
scaling method using as reference GRO J1655-40 D05 (dashed histogram
with positive slopes) and GX 339-4 D03 (dashed histogram
with negative slopes).
}
\label{fig4}
\end{figure} 
%%%----------------------------------------

We restricted our
analysis to the 2--10 keV energy range to avoid the complexity
related to the soft X-ray band, which includes the still debated nature of the soft
excess and the possible presence of warm absorbers. For completeness, we
tested whether and how the estimate of \mbh\ is affected by the use of the
hard X-ray data only. To this end, we first used \pks, a bright AGN for which we have
high quality proprietary \xmm\ data and simultaneous coverage of the optical/UV
band with the \swift\ UVOT (see \citealt{glioz10} for details). 
We performed a spectral fitting analysis of both
the extended optical/UV to X-ray range as well as 
of the restricted 2--10 keV energy band. The main
difference is the resulting temperature ($kT\simeq 8-20$ eV when using the
full range, as opposed to $kT\simeq 100$ eV obtained from the 2--10 keV band),
which however has a negligible impact on the estimate of \mbh. The resulting 
values of \mbh\ differ by less than 50\%. The little impact of
$kT$ (and of the parameter $\log(A)$ related to the Comptonization fraction)
on the \mbh\ estimate was further assessed considering the two AGNs of our sample
with the highest (Ark 120) and lowest (NCG 3227) Eddington ratio. When we 
arbitrarily varied by 80\% the best-fit values of $kT$ and $\log(A)$, the
consequent changes of \mbh\ were of the order of 20-30\% or less. We can 
therefore conclude that using the 2--10 keV range does not significantly
affect the determination of \mbh\ for this sample of AGNs. The results 
of the spectral analysis are summarized in Table 4.
%%--------------FIG5-------correlation BMC-RM FG=1-------------------------
\begin{figure*}
\begin{center}
\includegraphics[bb=100 85 550 760,clip=,angle=0,width=14cm]{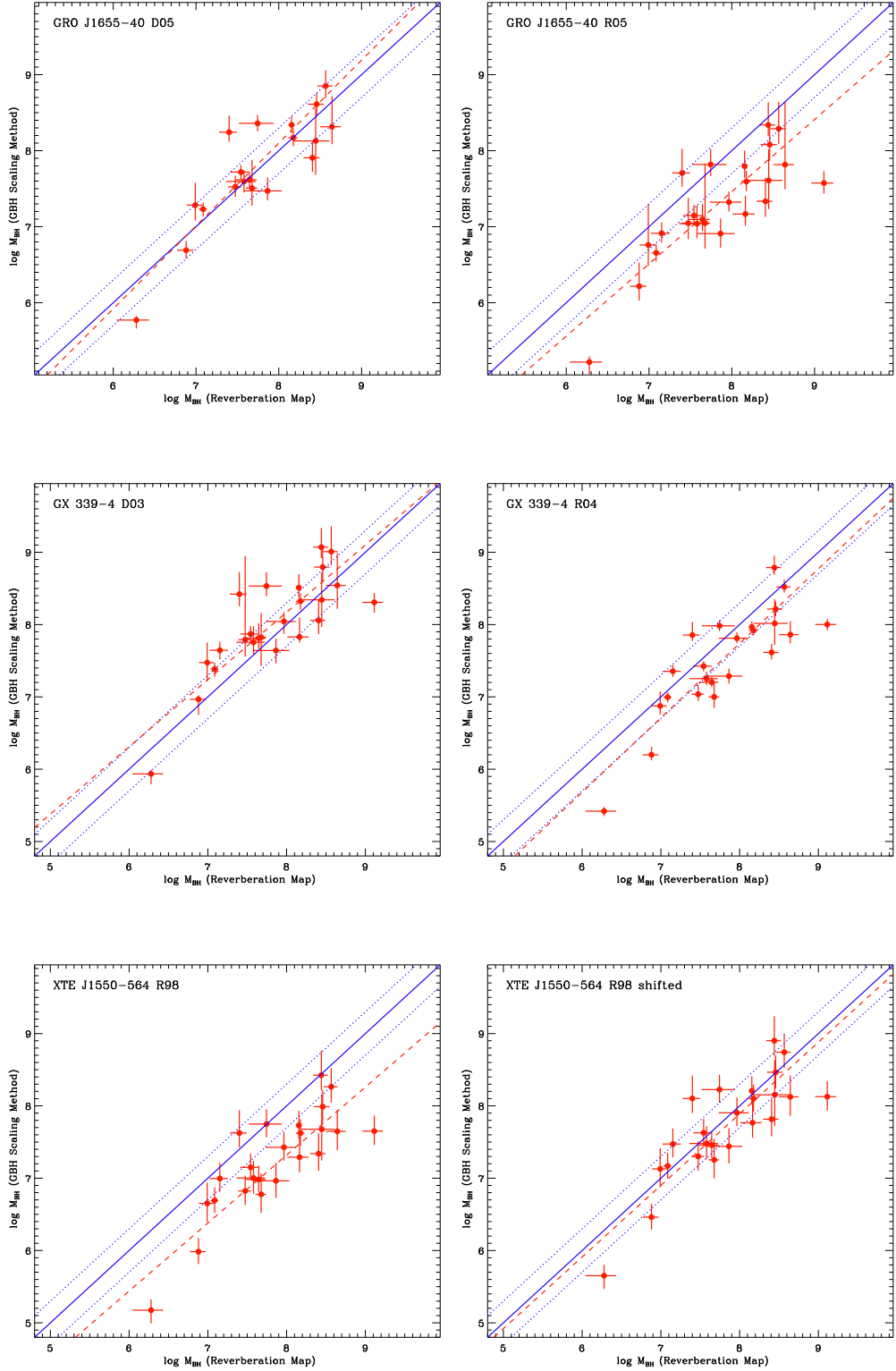}
\end{center}
\caption{\footnotesize
\mbh\ values obtained with the scaling method using different reference
sources plotted versus the reverberation mapping values. The dashed line 
is the linear best-fit, the 
thick solid line indicates the one-to-one
correlation; the dotted lines represent the 0.3 dex levels, commonly assumed as
uncertainty on the reverberation mapping estimates. The geometrical factor 
$F_{\rm G}$ is fixed to 1 illustrating a spherical geometry scenario. In the 
bottom-left panels the data points have been shifted along the y axis by a factor of 3. 
}
\label{fig5}
\end{figure*}
%%--------------FIG6-------correlation BMC-RM FG=FG(i)-------------------------
\begin{figure*}
\begin{center}
\includegraphics[bb=100 85 550 760,clip=,angle=0,width=14cm]{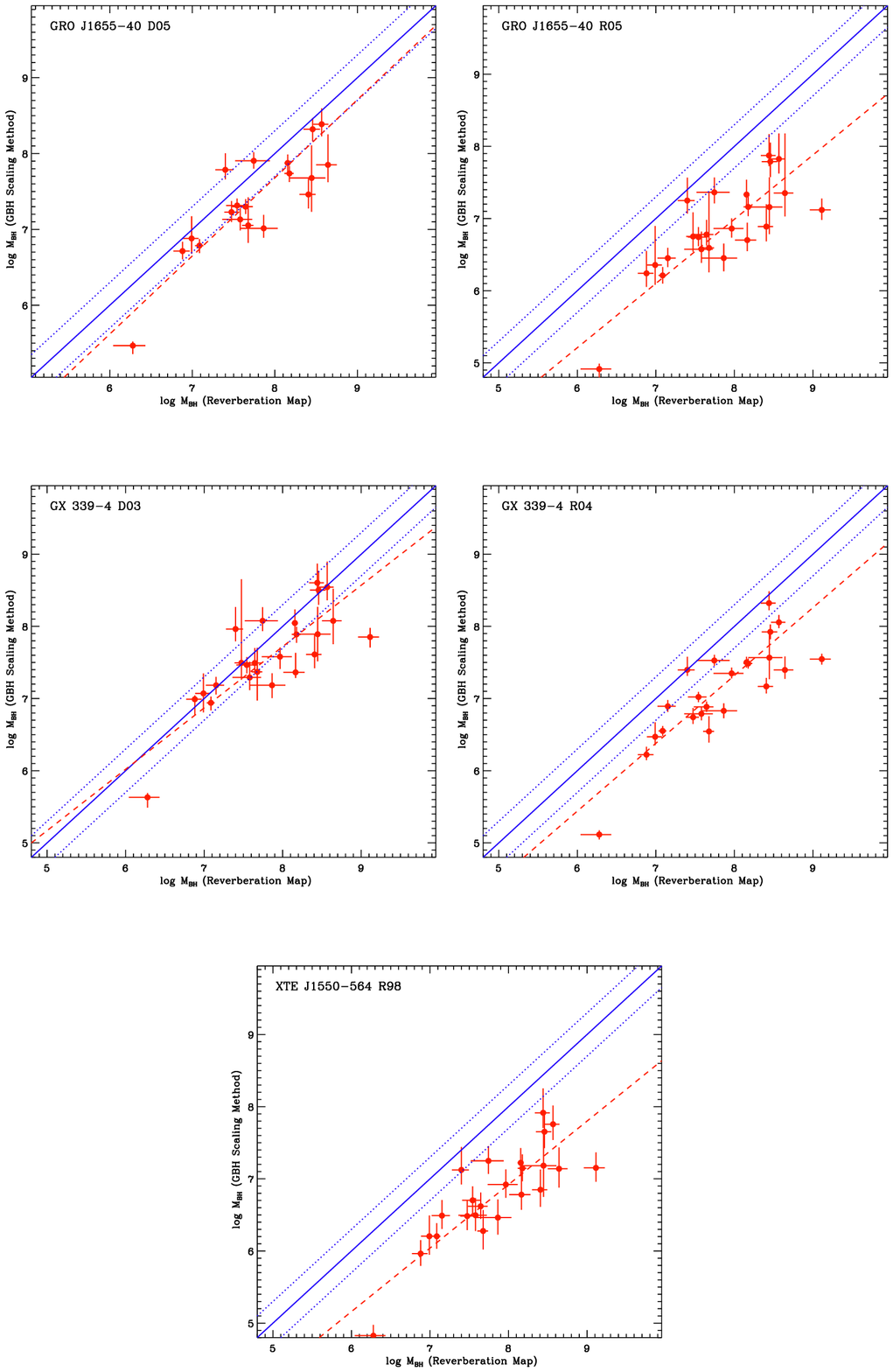}
\end{center}
\caption{\footnotesize
\mbh\ values obtained with the scaling method using different reference
sources plotted versus the reverberation mapping values. The dashed line
represents the linear best-fit, the solid line 
indicates the one-to-one correlation and the dashed lines are the 0.3 dex levels. 
The geometrical factor 
$F_{\rm G}$ is determined by the inclination angles reported in Tables 1 and 3
and illustrates a disk-like geometry.}
\label{fig6}
\end{figure*}

\subsection{Estimate of \mbh}
The estimate of \mbh\ with this scaling method crucially depends on the
properties (\mbh, distance, and perhaps inclination angle) of the GBH primary reference 
sources. Although there is a general agreement on the
GRO J1655-40 system parameters quoted in Table 1, it must be noted that 
the distance value has been questioned by \citet{foe06}, who proposed
a considerably smaller value ($d<$ 1.7 kpc). However, it was pointed out that
the systemic velocity reported by \citet{foe06} is at odds with
higher accuracy measurements reported in the literature \citep{gonz08} 
and that the lower value of the distance would not allow the companion
star to fill its Roche lobe \citep{caba07}. In addition, the fact
that using $d=$ 3.2 kpc ST09 obtain an excellent agreement between 
the values of \mbh\ 
determined with the scaling technique and the corresponding dynamical values  
 for several GBHs casts further doubts on the claim of \citet{foe06}.
In the remainder of the paper, we will therefore utilize for the reference sources
the system parameters provided in Table 1 (if we used $d=$ 1.7 kpc for GRO J1655-40
the AGN \mbh\ values would increase by a factor of $\sim$3.5).

For 24 sources, the BH mass was estimated using Equation (2)
and the five patterns of the reference sources shown in Fig.~\ref{fig1}. 
Five objects (Mrk 79, NGC 3516, NGC 4151, PG 1226+023, and PG 1411+442), perhaps
because of high intrinsic absorption and/or in reflection-dominated state,
yielded photon indices too low ($\Gamma < 1.4$) to allow a direct comparison
with the spectral trends of the reference sources introduced in Section 2.2.
In those cases, it was not possible to determine \mbh\ with this new scaling
method. The distribution of the logarithm of the Eddington ratio 
$\lambda_{\rm Edd}\equiv L_{\rm bol}/L_{\rm Edd}$ for remaining sample  
is shown in Figure~\ref{fig3}, which illustrates that $\lambda_{\rm Edd}$
spans a range of 0.01--1 with the majority of the
objects populating the 0.03-0.3 interval.

Since the geometry of the X-ray
emitting region is still a matter of debate, we have considered two different
scenarios: (1) spherical geometry, which is defined by setting $F_G=1$
in the \mbh\ equation, and (2) disk-like geometry where 
$f_G=\cos\theta_r/\cos\theta_t$ and the inclination angles given in Tables 1
and 3 are used for the reference sources (for GX 339-4 we used 70$^o$) and
the AGNs, respectively. 

The distributions of the AGN \mbh\ estimates obtained from the two 
most well-sampled reference transitions (GRO J1655 D05 and GX 339 D03) 
are shown in Figure~\ref{fig4} along with the distribution of values from 
the reverberation
mapping sample. The apparent general agreement between the distributions is formally
confirmed by a Kolmogorov-Smirnov (KS) test, which indicates that the 
distributions obtained with this new X-ray scaling method are indistinguishable
from the reverberation mapping one (for both reference sources the KS statistic
is $\sim$0.19 with an associated probability of $\sim$81\%). We also report the 
individual values of \mbh\ 
in Table 5 to allow a direct quantitative 
comparison with the corresponding values from reverberation mapping.
The errors quoted are
computed adding in quadrature the uncertainties on the \mbh\ and distance
of the reference sources (given in Table 1) and the uncertainty on the 
$(N_{\rm BMC,t}/N_{\rm BMC,r})$, which in turn depends on the 90\% confidence
errors of the X-ray spectral analysis on both $N_{\rm BMC}$ and $\Gamma$
(in Table 4)
as well as on the uncertainties of the functional parameters describing the
reference patterns in the $\Gamma - N_{\rm BMC}$ diagram.

The results in Table 5 suggest that this method provides \mbh\ estimates that
are well constrained (with an average percentage error of $\sim$45\% when using
the GRO J1655 D05 transition and $\sim$85\% for GX 339 D03)
and in general agreement with the corresponding values determined via 
reverberation mapping (the average ratio between the \mbh\ values determined with 
this method and those from reverberation mapping is $1.5\pm0.4$ for GRO J1655 D05
and $2.2\pm0.5$ for GX 339 D03).

Reasonable values for AGN \mbh\ are derived using all the reference spectral 
transitions available, as demonstrated by the values reported in
Table 6. The 4$th$ column of this table can be considered as the average departure
from the mass ratio of 1 and hence as a proxy of the
discrepancy between the values of \mbh\ determined with X-ray scaling 
method and the corresponding reverberation mapping ones.
More specifically, $\langle max(M_{\rm BH,ratio})\rangle$ is obtained by 
inverting the \mbh\ ratios that are lower than the unity (as an example,
for Fairall 9, instead of using 0.3 as reported in columns 3 of Table 5
we use 3.3=1/0.3) and then averaging over all \mbh\ ratios. In this way,
values much lower than unity will not cancel out with values much larger
than unity. The 3$^{rd}$ column of Table 6 reports the regular average ratio
between \mbh\ values determined with the two methods. This provides information 
about the tendency
of different reference patterns to either overestimate or underestimate \mbh\
with respect to the reverberation mapping values. For example, the spectral patterns corresponding to a decay phase of the outburst tend to slightly overestimate \mbh\
when a spherical geometry is assumed and to slightly underestimate it when a disk 
geometry is chosen. On the other hand, the rising patterns have a tendency
to underestimate \mbh\ by a factor of $\sim$2--3 when a spherical geometry is
considered and
by a larger amount when a disk-like geometry is used. 

We can conclude that all the reference patterns used with a 
quasi-spherical illuminating soft photon emission provide reliable  
estimates of \mbh\ in our AGN sample, with GRO J1655 D05 and XTE J1550 R98
showing the lowest and highest average discrepancy with respect to
the reverberation
mapping estimates ($\sim$2 and $\sim$6, respectively).

\subsection{Correlation Analysis}
An additional direct way to compare the \mbh\ estimates from this method with the 
corresponding values determined via reverberation mapping is to plot
one quantity versus the other and perform linear correlation tests.
The \mbh\ estimates derived from all the reference transitions are
plotted against the corresponding values obtained from the reverberation mapping 
in Figure~\ref{fig5} (spherical geometry) and Figure~\ref{fig6} 
(disk-like geometry), where the dashed line represents the linear best-fit,
the solid line indicates the one-to-one
correlation and the dotted lines are the 0.3 dex levels, commonly 
assumed as uncertainty on the reverberation mapping estimates. The 
vertical error-bars account for the uncertainties in the
BH mass and distance of the reference sources, for the uncertainty
of the functional reference and spectral parameters, 
as explained in the previous section. 

A visual inspection of these figures reveals a general agreement between the 
\mbh\ values obtained with these two completely independent techniques. This
consistency is more evident in the spherical geometry scenario shown in 
Figure~\ref{fig5}. In the bottom right panel of this figure, we show that
an excellent agreement can be obtained also using the XTE J1550-564 transition
as a reference, provided that the \mbh\ values are systematically increased 
by a factor of 3. The tight correlation observed in Figures~\ref{fig5}
and ~\ref{fig6} is formally
confirmed by non-parametric correlation analyses based on the Spearman's
$\rho$ and Kendalls's $\tau$ rank correlation coefficients 
as well as by a linear correlation fit using the 
routine \verb+fitexy+ \citep{press97} that accounts 
for the errors not only on the y-axis but along the x-axis as well.

In Table 7 we report the results of 
this analysis. From the results of the Spearman's
and Kendalls's analyses we infer that the \mbh\ values obtained with
these two different method are always correlated at high significance level
regardless of the reference transition used. This conclusion is further
supported by the fact that in all cases the best-fit line's 
slope is either consistent or very close to unity 
 (and inconsistent with the zero value at more than 10$\sigma$ level)
and the intercept is close to zero.

%%%%%%%%%%%%%%%%%%%%%%%%%%%%%%%%%%%%%%%%%%%%%%%%%%%%%%%%%%%%%%%%%%%%%%%%%%%%%%%%%
%%%%%%%%%%%%%%%%%%%%%%%%%%%% SECTION 5 CONCLUSION             %%%%%%%%%%%%%%%%%%%
%%%%%%%%%%%%%%%%%%%%%%%%%%%%%%%%%%%%%%%%%%%%%%%%%%%%%%%%%%%%%%%%%%%%%%%%%%%%%%%%%
\section{Discussion and Conclusions} 
In this work, we have explored whether a new X-ray scaling method 
recently introduced to determine the mass of stellar black holes
can be successfully extended to constrain the mass of supermassive black holes in AGNs. 
To this aim we have utilized a sample of  24 AGNs with $\Gamma=1.56-2.11$
and $\lambda_{\rm Edd}=0.01-1$, and
with \mbh\ previously determined 
via reverberation mapping and with good-quality archival \xmm\ data.
The main results can be summarized as follows.

\begin{itemize}

\item  This novel method, which is based on the spectral fitting
of X-ray data with the BMC Comptonization model and on the similarity
of the X-ray spectral behavior in BHs regardless of of their mass, 
appears to be a robust estimator of \mbh\ in AGNs. 

\item The values of \mbh\ determined with this method are well constrained
and in good agreement (within a factor of $\sim$2--3)
with the corresponding values obtained with the
reverberation mapping technique. 

\item For AGNs with $\Gamma$ comprised in the 1.4-1.95 range
the most reliable reference appears to be the outburst decay pattern 
of GRO J1655-40 in combination with spherical scenario (i.e., with
$F_G=1$ in Eq. 2). For $\Gamma\le 2.1$
both decay and rise patterns of GX 339-4 and spherical geometry provide
reliable estimates of \mbh. Finally, for steeper photon indices up to 
$\Gamma\sim3$ only the rising pattern of XTE J1550-564 can be used,
keeping in mind its tendency to underestimate \mbh\ by a factor of
$\sim$3.

%\item  An excellent agreement is also found using XTE J1550-564 as a 
%reference and multiplying the results by a factor of 3. The large range
%of photon indices spanned by XTE J1550-564 during its spectral transition 
%is of crucial importance to estimate \mbh\ in AGNs with steep $\Gamma$.

\item The \mbh\ values appear to be systematically
underestimated by a factor of 2-3 when a disk-like geometry is used
instead of a quasi-spherical geometry for the soft photon supply. This result
lends support to the hypothesis that the soft photon emission
area is quasi-spherical.

\end{itemize}

In summary, our analysis reveals that this X-ray scaling technique yields 
black hole masses that are in full agreement with the accepted values 
obtained from the reverberation mapping. This is consistent
with the hypothesis that 
the X-ray spectrum  produced by the Comptonization process is self-similar 
and its shape is independent of \mbh, and lends further support to 
grand unification model between AGNs and GBHs.

Before concluding it is important to briefly discuss the range of 
applicability of this technique and its main differences with respect 
to most commonly used methods for black hole mass determination. 
First, this is one of the few truly scale-independent techniques 
to estimate \mbh. The other two scale-independent methods are the one based
on the so called ``fundamental plane" of black
holes (BHs) introduced by \citet{merlo03} and  \citet{falc04}, 
where $M_{\rm BH}$ is related to both the X-ray and radio luminosities, 
and the one based
on the scaling of temporal breaks in X-ray power spectral density analyses
\citep{mchar06}. However, the fundamental plane is affected by large scattering
and is limited to BH systems characterized by radio emission, whereas the
method based on X-ray timing
requires high-quality evenly-sampled X-ray light curves spanning a baseline of 
more than a decade and therefore is necessarily limited to very few selected 
AGNs. On the other hand, this X-ray scaling novel technique
 only needs moderately good X-ray spectra, which are now
available for a very large number of AGNs and BH systems in general.
Second, unlike most indirect methods, this X-ray scaling method
is completely independent of any assumption  on the BLR nature/geometry or 
host galaxy characteristics, and can therefore complement several 
optically-based studies
by providing independent estimates of \mbh, and expand them when the
optical properties are uncertain or unavailable.

In principle, this technique can be applied to 
any accreting black hole with bulk X-ray emission produced via Comptonization
and for which the photon index falls in a range covered by the GBH reference
patterns in the $\Gamma - N_{\rm BMC}$ diagram. However, we caution that a blind
application of this technique may lead to erroneous results. This will occur
whenever the genuine properties of the primary Comptonized X-ray emission
are not properly determined, for example because 
effects associated with strong absorption, reflection, or emission from an
additional component such as a relativistic jet are not carefully accounted
for. Particular care should be used when dealing with low accreting systems,
because this method has not been tested below $\lambda_{\rm Edd}=0.01$
and because at very low values of $\dot m$ a radiatively inefficient accretion 
mode (possibly accompanied by outflows and jets)
is likely to take place leading to the bulk of the X-rays to be dominated 
by jet emission as suggested by \citet{yuancui05}.

It is also worth mentioning that AGNs are generally characterized by high flux 
variability, which is often accompanied by spectral variability. The vast
majority of bright radio-quiet AGNs show steeper spectra when they 
are brighter and 
hence their evolution is in qualitative agreement with the rising trend in the 
$\Gamma - N_{\rm BMC}$ plot, providing strong support for the physical basis 
of our methodology. Additionally, AGN flux and spectral changes are
generally not so extreme and therefore the value obtained with different 
observations will be within the uncertainty of the method. Finally, occurrences
of extreme spectral changes are usually associated with changing in the local
obscuration component (for example from Compton-thick to Compton-thin scenario)
and therefore do not reflect the intrinsic change of the Comptonized primary component. A possible notable exception is NGC 4051, which, in virtue of its
small SMBH, shows extreme flux and spectral changes that might reflect
genuine changes of the primary X-ray component. This is indeed the only 
source for which a complete trend in the $\Gamma - N_{\rm BMC}$ plot can be
constructed. Preliminary results based on numerous \xmm\ and \chandra\
observations suggest that NGC 4051 follow a trend consistent with
GBH references and that the scaling method successfully determine its mass
(Chekhtman \& Titarchuk 2011, in preparation).

In the near future,  we plan to systematically apply this 
X-ray scaling method to different categories
of objects, whose \mbh\ values are currently debated, such as
ultraluminous X-ray sources, narrow line Seyfert 1 galaxies,
non-jet-dominated radio loud AGNs, and powerful quasars
to test the upper end of the \mbh\ -- $\sigma_\star$ relationship.

\begin{acknowledgments} 
We thank the anonymous referee for constructive suggestions that have improved
the clarity of the paper.
We thank Nikolai Shaposhnikov for providing the spectral data of XTE J1550-564
and for useful discussions. MG acknowledges support by the NASA ADP grant NNXlOAD51G. 
\end{acknowledgments}

%\bibliographystyle{plainnat}

%%%%%%%%%%%%%%%%%%% TAB1: REFERENCE SOURCES %%%%%%%%%%%%%%%%%%%%%%%%%%%%%
\begin{table} 
\footnotesize
\caption{Characteristics of reference sources}
\begin{center}
\begin{tabular}{cccc} 
\hline        
\hline
\noalign{\smallskip}
Name    &  $M_{\rm BH}$  & $d$   & $i$ \\
        &  ($M_{\odot}$) & (kpc) &  $(\deg)$\\
(1)           & (2)   & (3)   & (4)         \\
\noalign{\smallskip}
\hline 
\noalign{\smallskip}
GRO J1655-40      &    $6.3\pm0.3$  &  $3.2\pm0.2$  & $70\pm1$ \\
\noalign{\smallskip}
GX 339-4     &    $12.3\pm1.4$  &  $5.7\pm0.8$  & $>45$ \\
\noalign{\smallskip}
XTE J1550-564      &    $10.7\pm1.5$  &  $3.3\pm0.5$  & $72\pm5$ \\
\noalign{\smallskip}
\hline        
\hline
\end{tabular}
\end{center}
{\bf Columns Table 1}: 1= Source name. 2= Black hole mass. 
3= Distance. 4= Inclination angle. {\bf Notes:} For GRO J1655-40
the quoted values of $M_{\rm BH}$, $d$, and  $i$ are from \citet{gree01}
and \citet{hjel95}. For both GX 339-4
and XTE J1550-564 $M_{\rm BH}$ and $d$ are from ST09, whereas 
the values of $i$ are respectively from from \citet{kole10} and from \citet{oro02}.  
\label{tab1}
\footnotesize
\end{table}  
%%%%%%%%%%%%%%%%%%% TAB2: GAMMA-N_BMC PARAMETRIZATION OF REFERENCE SOURCES %%%%%%%%%%%%%%%%
\begin{table*}[htb] 
\footnotesize
\caption{Parametrization of $\Gamma-N_{\rm BMC}$ reference patterns}
\begin{center}
\begin{tabular}{cccccc} 
\hline        
\hline
\noalign{\smallskip}
Transition    &  $A$  & $B$   & $N_{\rm tr}$ & $\beta$ & $\chi^2_r$ (dof)\\
(1)           & (2)   & (3)   & (4)          & (5)     & (6)\\
\noalign{\smallskip}
\hline 
\noalign{\smallskip}
GRO J1655 D05  & $1.96\pm0.02$  &  $0.42\pm0.02$  & $0.023\pm0.001$ &  $1.8\pm0.2$  & 1.5 (20)\\
\noalign{\smallskip}
GRO J1655 R05  & $2.35\pm0.04$  &  $0.74\pm0.04$  & $0.131\pm0.001$ &  $1.0\pm0.1$  & 1.8 (34)\\
\noalign{\smallskip}
GX339 D03   &  $2.13\pm0.03$  &  $0.50\pm0.04$  & $0.0130\pm0.0002$ &  $1.5\pm0.3$  & 0.9 (40)\\
\noalign{\smallskip}
GX339 R04   &  $2.10\pm0.03$  &  $0.46\pm0.01$  & $0.037\pm0.001$ &  $8.0\pm1.5$  & 2.3 (24)\\
\noalign{\smallskip}
XTE J1550 R98  & $2.96\pm0.02$  &  $2.8\pm0.2$  & $0.055\pm0.010$ &  $0.4\pm0.1$  & 2.7 (49)   \\
\noalign{\smallskip}
\hline  
\hline   
\end{tabular}
\end{center}
{\bf Columns Table 2}: 1= Reference source spectral transition. 2= Parameter
of the functional form described in Eq.(1) that is responsible for the rigid
translation of the spectral pattern along the y-axis. 3= Parameter 
characterizing the lower saturation level of the pattern. 4= Parameter 
responsible for the translation of the spectral pattern along the x-axis.
5= Slope of the spectral pattern. 6= Reduced $\chi^2$ and degrees of freedom.
{\bf Notes}: The quoted parameter errors are the 1$\sigma$ uncertainties obtained from
fitting the spectral data using the Levenberg-Marquardt algorithm.
\label{tab2}
\footnotesize
\end{table*}  
%%%%%%%%%%%%%%%%%%% TAB3: AGN SAMPLE PROPERTIES %%%%%%%%%%%%%%%%
\begin{table}[htb] 
\footnotesize
\caption{Sample properties}
\begin{center}
\begin{tabular}{ccccccc} 
\hline        
\hline
\noalign{\smallskip}
Name    &       $z$  &&   $(i)$  &  $\log(M_{\rm BH})$ & $\log(L_{\rm bol})$   & $\lambda_{\rm Edd}$ \\
        &                &&    (deg)     &  ($M_{\odot}$)   &  (${\rm erg~s^{-1}}$) &  $(L_{\rm bol}/L_{\rm Edd})$\\
(1)   &     (2)          && (3)       & (4)                 &(5)                    & (6)\\
\noalign{\smallskip}
\hline 
\noalign{\smallskip}
3C 120      &     0.0330 &&  $13_{-6}^{+11}$ &       7.74              &    45.3         & 0.305 \\
\noalign{\smallskip}
3C 390.3    &     0.0561 &&  $48_{-20}^{+24}$  &       8.46              &    45.2         & 0.047 \\
\noalign{\smallskip}
Ark 120     &     0.0327 &&  $22_{-9}^{+16}$ &       8.18              &    45.3         & 0.111  \\
\noalign{\smallskip}
Fairall 9   &     0.0470 &&  $17_{-7}^{+13}$  &       8.41              &    44.8         & 0.019 \\
\noalign{\smallskip}
Mrk 110     &     0.0353 &&  $11_{-5}^{+9}$  &       7.40              &    45.1         & 0.433  \\
\noalign{\smallskip}
Mrk 279     &     0.0305 &&   30 &       7.54              &    45.0         & 0.210  \\
\noalign{\smallskip}
Mrk 335     &     0.0258 &&  $8_{-4}^{+6}$  &       7.15              &    45.3         & 1.130  \\
\noalign{\smallskip}
Mrk 509     &     0.0344 &&  $5_{-2}^{+4}$   &       8.16              &    45.2         & 0.095 \\
\noalign{\smallskip}
Mrk 590     &     0.0264 &&  $13_{-6}^{+9}$&       7.68              &    43.8         & 0.010 \\
\noalign{\smallskip}
Mrk 79      &     0.0222 &&  $31_{-13}^{+23}$  &       7.72              &    44.3         & 0.031 \\
\noalign{\smallskip}
NGC 3227    &     0.0039 &&  $71_{-24}^{+17}$ &       6.88              &    42.9         & 0.001 \\
\noalign{\smallskip}
NGC 3516    &     0.0088 &&  30 &       7.50              &    43.5         & 0.006 \\
\noalign{\smallskip}
NGC 3783    &     0.0097 &&  $48_{-21}^{+27}$ &       7.47              &    44.2         & 0.043 \\
\noalign{\smallskip}
NGC 4051    &     0.0023 &&  $46_{-20}^{+27}$ &       6.23              &    42.6         & 0.016 \\
\noalign{\smallskip}
NGC 4151    &     0.0033 &&  $77_{-22}^{+10}$  &       7.12              &    44.0         & 0.056 \\
\noalign{\smallskip}
NGC 4593    &     0.0090 &&   30 &       6.99              &    43.7         & 0.037 \\
\noalign{\smallskip}
NGC 5548    &     0.0172 &&  $45_{-18}^{+23}$   &       7.65              &    44.3         & 0.024 \\
\noalign{\smallskip}
NGC 7469    &     0.0163 &&  $18_{-8}^{+13}$   &       7.09              &    44.8         & 0.369 \\
\noalign{\smallskip}
PG 0052+251 &     0.1550 &&  $5_{-2}^{+4}$  &       8.57              &    45.8         & 0.148 \\
\noalign{\smallskip}
PG 0844+349 &     0.0640 &&  $8_{-3}^{+6}$ &       7.97              &    45.4         & 0.233 \\
\noalign{\smallskip}
PG 0953+414 &     0.2341 &&  $2_{-1}^{+2}$    &       8.44              &    46.5         & 0.892 \\
\noalign{\smallskip}
PG 1211+143 &     0.0809 &&  $3_{-1}^{+2}$  &       8.16              &    45.7         & 0.260 \\
\noalign{\smallskip}
PG 1226+023 &     0.1583 &&  $1_{-1}^{+1}$  &       8.95              &    47.1         & 1.140 \\
\noalign{\smallskip}
PG 1229+204 &     0.0630 & & $12_{-5}^{+10}$  &       7.86              &    44.9         & 0.082 \\
\noalign{\smallskip}
PG 1307+085 &     0.1550 &&  $5_{-2}^{+5}$  &       8.64              &    45.6         & 0.066 \\   
\noalign{\smallskip}
PG 1411+442 &     0.0896 &&  $3_{-1}^{+2}$  &       8.65              &    45.4         & 0.041 \\ 
\noalign{\smallskip}
PG 1426+015 &     0.0865 &&  $13_{-6}^{+11}$  &       9.11              &    45.6         & 0.024 \\              
\noalign{\smallskip}
PG 1613+658 &     0.1290 &&  $15_{-7}^{+12}$  &       8.45              &    45.9         & 0.221 \\        
\noalign{\smallskip}
PG 2130+099 &     0.0630 &&  $6_{-3}^{+4}$  &       7.58              &    45.0         & 0.018 \\
\noalign{\smallskip}
\hline
\hline
\end{tabular}
\end{center}
{\bf Columns Table 3}: 1= Source name. 2= Redshift. 3= Inclination angle. 
4= Black hole mass determined via reverberation mapping. 
5= Bolometric luminosity. 6= Eddington ratio. 
{\bf Notes}: The AGN distances were computed from $z$ assuming 
$H_0=71~{\rm km~s^{-1}~Mpc^{-1}}$, $\Omega_\Lambda=0.73$, and 
$\Omega_m=0.27$ \citep{sper03}. For nearby AGNs we used 
the redshift-independent distances provided by the NASA/IPAC extragalactic  
database (NED). The quoted values of the inclination angle are from
\citet{bian02}; for objects without estimate of inclination angle
(Mrk 279, NGC 3516, NGC 4593) we assumed a value of 30$^o$ typical for
Seyfert galaxies. \mbh\ values are from \citet{peter04} with the
exception of NGC 3227, NGC 3516, NGC 4051, NGC 5548 \citep{denn10},
and PG 2130+099 \citep{grier08}. Values for  $(L_{\rm bol})$ and 
$\lambda_{\rm Edd}$ are from \citet{vasu09}.
\label{tab3}
\footnotesize
\end{table}

%%%%%%%%%%%%%%%%%%% TAB4: AGN X-RAY SPECTRAL PROPERTIES %%%%%%%%%%%%%%%% 
\begin{table*}[htb] 
\footnotesize
\caption{Spectra Results}
\begin{center}
\begin{tabular}{cccccccc} 
\hline        
\hline
\noalign{\smallskip}
Name & XMM obsid & Net exposure &$kT$  & $Log(A)$ &  $\Gamma$ & $N_{\rm BMC}$  & $ \chi_{r}^2$(dof)\\
     &           &  (ks)        & (keV) &          &           & ($10^{-4}$) &                   \\
(1)  & (2)       & (3)          & (4)      &  (5)      & (6)         & (7)    & (8)  \\
\noalign{\smallskip}
\hline 
\noalign{\smallskip}
3C 120 &  0152840101 & 79.1 & $0.4\pm0.1$ & $0.7\pm0.1$    & $1.69_{-0.02}^{+0.02}$ & $4.5_{-0.1}^{+0.1}$ & 1.02(1211)\\
\noalign{\smallskip}
3C 390.3 & 0203720201 & 35.0  & $0.1\pm0.1$ & 0.7          & $1.66_{-0.03}^{+0.04}$ & $2.5_{-0.2}^{+0.2}$ & 1.13(817)\\
\noalign{\smallskip}
Ark 120 & 0147190101 & 55.9 & $0.4\pm0.4$ & $0.5\pm0.1$   & $1.84_{-0.03}^{+0.03}$ & $4.2_{-0.1}^{+0.1}$ & 1.07(1000)\\
\noalign{\smallskip}
Fairall 9 & 0101040201 & 26.0  & $0.1\pm0.1$ & $0.5\pm0.1$ & $1.80_{-0.10}^{+0.07}$ & $1.0_{-0.1}^{+0.1}$ & 1.05(866)\\
\noalign{\smallskip}
Mrk 110 & 0201130501 & 32.9 & $0.3\pm0.1$ & $0.6\pm0.3$   & $1.68_{-0.02}^{+0.03}$ & $2.7_{-0.1}^{+1.0}$ & 0.99(1290)\\
\noalign{\smallskip}
Mrk 279 & 0302480501& 34.9 & 0.05 & 0.95                   & $1.80_{-0.01}^{+0.01}$ & $1.5_{-0.1}^{+0.1}$ & 1.05(1252)\\
\noalign{\smallskip}
Mrk 335 & 0306870101 & 83.3 &  $0.3\pm0.1$ & $0.4\pm0.1$   & $1.98_{-0.03}^{+0.03}$ & $2.2_{-0.1}^{+0.2}$ & 0.99(1364)\\
\noalign{\smallskip}
Mrk 509 & 0306090401 & 44.9 & $0.3\pm0.1$ & $0.6\pm0.1$   & $1.69_{-0.02}^{+0.02}$ & $4.1_{-0.1}^{+0.1}$ & 1.02(1468)\\
\noalign{\smallskip}
Mrk 590 & 0201020201 & 39.5 & $0.4\pm0.1$ & $0.8\pm0.2$   & $1.54_{-0.11}^{+0.05}$ & $0.6_{-0.1}^{+0.1}$ & 1.03(838)\\
\noalign{\smallskip}
Mrk 79 & 0502091001 & 49.0 & 0.11 & 0.7                     & $1.22_{-0.02}^{+0.02}$ & $0.4_{-0.1}^{+0.1}$ & 1.02(1065)\\
\noalign{\smallskip}
NGC 3227 & 0400270101 & 96.1 & 0.08 & 0.5                   & $1.56_{-0.003}^{+0.003}$ & $2.9_{-0.1}^{+0.1}$ & 0.98(1574)\\
\noalign{\smallskip}
NGC 3516 & 0107460701 & 81.8 & $0.2\pm0.1$ & 0.5          & $1.32_{-0.01}^{+0.02}$ & $1.3_{-0.1}^{+0.1}$ & 1.06(1411)\\
\noalign{\smallskip}
NGC 3783 & 0112210501 & 89.5 & $0.2\pm0.1$ & 0.8          & $1.56_{-0.01}^{+0.01}$ & $4.2_{-0.2}^{+0.2}$ & 1.19(1325)\\
\noalign{\smallskip}
NGC 4051 & 0606320101 & 45.7 &  $0.1\pm0.1$ & $0.1\pm0.1$  & $1.72_{-0.004}^{+0.004}$ & $1.5_{-0.1}^{+0.1}$ & 0.99(1059)\\
\noalign{\smallskip}
NGC 4151 & 0143500301 & 12.4 & 0.01 & 2.0                  & $1.35_{-0.01}^{+0.01}$ & $3.4_{-0.1}^{+0.1}$ & 1.30(1593)\\
\noalign{\smallskip}
NGC 4593 & 0109970101 & 5.0 & $0.4\pm0.1$ & $0.6\pm0.2$  & $1.65_{-0.09}^{+0.08}$ & $3.6_{-0.4}^{+0.4}$ & 0.94(648)\\
\noalign{\smallskip}
NGC 5548 & 0089960301 & 50.5 & 0.10 & $0.6\pm0.1$           & $1.64_{-0.01}^{+0.01}$ & $2.5_{-0.1}^{+0.1}$ & 1.01(1494)\\
\noalign{\smallskip}
NGC 7469 & 0207090101 & 58.3 & $0.2\pm0.1$& 1.2           & $1.86_{-0.01}^{+0.01}$ & $2.0_{-0.1}^{+0.1}$ & 1.15(902)\\
\noalign{\smallskip}
PG 0052+251 & 0301450401 & 8.2 & $0.3\pm0.2$ & 0.5       & $1.75_{-0.11}^{+0.08}$ & $0.6_{-.1}^{+0.1}$ & 0.93(243)\\
\noalign{\smallskip}
PG 0844+349 & 0103660201 & 12.7 & $0.3\pm0.1$& $0.4\pm0.4$ & $2.00_{-0.14}^{+0.12}$ & $0.6_{-0.1}^{+0.9}$ & 0.96(297)\\
\noalign{\smallskip}
PG 0953+414 & 0111290201 & 11.2 & $0.1\pm0.1$ & 0.1         & $2.00_{-0.11}^{+0.07}$ & $0.5_{-0.1}^{+0.1}$ & 0.74(170)\\
\noalign{\smallskip}
PG 1211+143 & 0502050101 & 45.6 & $0.1\pm0.1$ & 2.0         & $2.11_{-0.03}^{+0.03}$ & $0.4_{-0.1}^{+0.2}$ & 0.90(659)\\
\noalign{\smallskip}
PG 1226+023 & 0414190101 & 45.7 & 0.10 & 0.5                  & $1.39_{-0.01}^{+0.01}$ & $6.4_{-0.1}^{+0.1}$ & 1.04(1595)\\
\noalign{\smallskip}
PG 1229+204 & 0301450201 & 16.9 & 0.05 & 1.0                  & $1.91_{-0.06}^{+0.06}$ & $0.3_{-0.1}^{+0.1}$ & 0.94(253)\\
\noalign{\smallskip}
PG 1307+085 & 0110950401 & 10.6 & 0.10 & 0.6                  & $1.60_{-0.09}^{+0.10}$ & $0.1_{-0.1}^{+0.1}$ & 0.91(98)\\
\noalign{\smallskip}
PG 1411+442 & 0103660101 & 22.4 & 0.07 & 2.0                  & $1.00_{-1.00}^{+1.00}$ & $3.4_{-0.3}^{+0.3}$ & 0.92(41)\\
\noalign{\smallskip}
PG 1426+015 & 0102040501 & 5.4 & 0.03 & 1.2                  & $1.97_{-0.06}^{+0.06}$ & $0.4_{-0.1}^{+0.1}$ & 0.88(214)\\
\noalign{\smallskip}
PG 1613+658 & 0102040601 & 5.0 & 0.04 & 1.2                  & $1.94_{-0.26}^{+0.29}$ & $0.3_{-0.1}^{+0.3}$ & 1.05(14)\\
\noalign{\smallskip}
PG 2130+099 & 0150470701 & 21.9 & 0.06   & 0.7                & $1.73_{-0.05}^{+0.05}$ & $0.2_{-0.1}^{+0.1}$ & 0.98(339)\\
\noalign{\smallskip}
\hline
\hline
\end{tabular}
\end{center}
\label{tab4}
{\bf Columns Table 4}: 1= Source name. 2= \xmm\ observation identifier.
3= EPIC pn net exposure time in ks. 4= Temperature of thermal photon source 
in keV. 5= Logarithm of the $A$ parameter, which is related to the Comptonization 
fraction $f$ by the relation $f=A/(1+A)$. 6= 2--10 keV photon index. 
7= Normalization of the Comptonization model in units of $(L/10^{39}{~\rm erg~s^{-1}})
(10~{\rm kpc}/d)^2$. 8= Reduced $\chi^2$ and degrees of freedom. {\bf Notes}:
In the cases, where the 
\verb+error+ procedure in {\tt XSPEC} was not able to provide the 90\% confidence 
uncertainty for $kT$ or $\log(A)$ due to low S/N data or spectral degeneracy,
we fixed the parameters at their best-fit value. The upper limit of $\log(A)$
was fixed at the value of 2.
\end{table*}

%%%%%%%%%%%%%%%%%%% TAB5: MBH USING PHASE DECAY TRANSITIONS %%%%%%%%%%%%%%%%
\begin{table*} 
\footnotesize
\caption{Black hole mass estimates}
\begin{center}
\begin{tabular}{ccccccccc} 
\hline        
\hline
\noalign{\smallskip}
Name &  \multicolumn{4}{c}{GRO J1655 D05} & \multicolumn{4}{c}{GX 339 D03} \\
     &\multicolumn{2}{c}{Sphere} & \multicolumn{2}{c}{Disk} & \multicolumn{2}{c}{Sphere} & \multicolumn{2}{c}{Disk}\\
     & $\log M_{\rm BH}$ & scal/RM &  $\log M_{\rm BH}$ & scal/RM&
        $\log M_{\rm BH}$ & scal/RM &  $\log M_{\rm BH}$ & scal/RM\\
(1) & (2) & (3) & (4) & (5) & (6) & (7) & (8) & (9)\\
\noalign{\smallskip}
\hline 
\noalign{\smallskip}
3C 120 &  $8.36\pm0.17$ & 4.1 & $7.91\pm0.17$ & 1.4 & $8.53\pm0.21$ & 6.1 & $8.08\pm0.21$ & 2.2\\
\noalign{\smallskip}
3C 390.3 &  $8.61\pm0.20$ & 1.4 & $8.32\pm0.20$ & 0.7 & $8.80\pm0.27$ & 2.2 & $8.50\pm0.27$ & 1.1\\
\noalign{\smallskip}
Ark 120  & $8.17\pm0.16$ & 1.0 & $7.74\pm0.16$ & 0.4 & $8.32\pm0.17$ & 1.4 & $7.89\pm0.17$ & 0.5\\
\noalign{\smallskip}
Fairall 9 & $7.91\pm0.24$ & 0.3 & $7.46\pm0.24$ & 0.1 & $8.06\pm0.27$ & 0.5 & $7.61\pm0.27$ & 0.2\\
\noalign{\smallskip}
Mrk 110 & $8.24\pm0.29$ & 7.0 & $7.78\pm0.29$ & 2.4 & $8.42\pm0.28$ & 10.5 & $7.96\pm0.28$ & 3.7\\
\noalign{\smallskip}
Mrk 279	& $7.71\pm0.25$ & 1.5 & $7.32\pm0.25$ & 0.6 & $7.87\pm0.17$ & 2.1 & $7.47\pm0.17$ & 0.8\\
\noalign{\smallskip}
Mrk 335 & $\dots$ & $\dots$ & $\dots$ & $\dots$ & $7.65\pm0.18$ & 3.1 & $7.18\pm0.18$ & 1.1\\
\noalign{\smallskip}
Mrk 509 & $8.34\pm0.17$ & 1.5 & $7.89\pm0.17$ & 0.5 & $8.51\pm0.21$ & 2.3 & $8.05\pm0.21$ & 0.8\\
\noalign{\smallskip}
Mrk 590 & $7.51\pm0.34$ & 0.7 & $7.04\pm0.34$ & 0.2 & $7.82\pm0.40$ & 1.4 & $7.37\pm0.40$ & 0.5\\
\noalign{\smallskip}
NGC 3227 & $6.69\pm0.17$ & 0.6 & $6.24\pm0.17$ & 0.2 & $6.97\pm0.20$ & 1.2 & $6.99\pm0.20$ & 1.3\\
\noalign{\smallskip}
NGC 3783 & $7.52\pm0.19$ & 1.1 & $7.55\pm0.19$ & 1.2 & $7.79\pm0.73$ & 2.1 & $7.50\pm0.73$ & 1.1\\
\noalign{\smallskip}
NGC 4051 & $5.77\pm0.15$ & 0.3 & $5.48\pm0.15$ & 0.2 & $5.94\pm0.17$ & 0.5 & $5.63\pm0.17$ & 0.2\\
\noalign{\smallskip}
NGC 4593 & $7.28\pm0.28$ & 2.0 & $6.98\pm0.28$ & 1.0 & $7.47\pm0.31$ & 3.0 & $7.07\pm0.31$ & 1.2\\
\noalign{\smallskip}
NGC 5548 & $7.62\pm0.16$ & 0.9 & $7.21\pm0.16$ & 0.4 & $7.81\pm0.22$ & 1.5 & $7.49\pm0.22$ & 0.7\\
\noalign{\smallskip}
NGC 7469 & $7.23\pm0.16$ & 1.4 & $6.91\pm0.16$ & 0.7 & $7.39\pm0.16$ & 2.0 & $6.94\pm0.16$ & 0.7\\
\noalign{\smallskip}
PG 0052+251 & $8.85\pm0.23$ & 1.9 & $8.41\pm0.23$ & 0.7 & $9.01\pm0.31$ & 2.8 & $8.54\pm0.31$ & 1.0\\
\noalign{\smallskip}
PG 0844+349 & $\dots$ & $\dots$ & $\dots$ & $\dots$ & $8.04\pm0.20$ & 1.2 & $7.58\pm0.20$ & 0.4\\
\noalign{\smallskip}
PG 0953+414 & $\dots$ &$\dots$ & $\dots$ & $\dots$ & $9.07\pm0.25$ & 4.3 & $8.61\pm0.25$ & 1.5\\
\noalign{\smallskip}
PG 1211+143 & $\dots$ &$\dots$ & $\dots$ & $\dots$ & $7.83\pm0.22$ & 0.5 & $7.36\pm0.22$ & 0.2\\
\noalign{\smallskip}
PG 1229+204 & $7.47\pm0.20$ & 0.4 & $7.01\pm0.20$ & 0.1 & $7.64\pm0.22$ & 0.6 & $7.18\pm0.22$ & 0.2\\
\noalign{\smallskip}
PG 1307+085 & $8.32\pm0.35$ & 0.5 & $7.85\pm0.35$ & 0.2 & $8.54\pm0.42$ & 0.8 & $8.08\pm0.42$ & 0.3\\
\noalign{\smallskip}
PG 1426+015 & $\dots$ & $\dots$ & $\dots$ & $\dots$ & $8.31\pm0.19$ & 0.2 & $7.85\pm0.19$ & 0.1\\
\noalign{\smallskip}
PG 1613+658 & $8.13\pm0.48$ & 0.5 & $7.66\pm0.48$ & 0.2 & $8.34\pm0.41$ & 0.8 & $7.89\pm0.41$ & 0.3\\
\noalign{\smallskip}
PG 2130+099 & $7.60\pm0.20$ & 1.0 & $7.58\pm0.20$ & 0.4 & $7.76\pm0.24$ & 1.5 & $7.29\pm0.24$ & 0.5\\
\hline  
\hline
\end{tabular}
\end{center}
{\bf Columns Table 5}: 1= Source Name. 2= \mbh\ in solar units computed using 
the GRO J1655 D05 transition in the case of spherical geometry. 3= Ratio
between the \mbh\ in column 2 and the corresponding value determined via
reverberation mapping. 4= Same as column 2 but for a disk-like geometry.
5= Ratio between the \mbh\ in column 3 and the corresponding value 
determined via reverberation mapping. 6= \mbh\ in solar units computed 
using the GX 339 D03 transition in the case of spherical geometry.  7= Ratio
between the \mbh\ in column 6 and the corresponding value determined via
reverberation mapping. 8= Same as column 6 but for a disk-like geometry.
9=  Ratio between the \mbh\ in column 8 and the corresponding value 
determined via reverberation mapping.
{\bf Notes}: For five sources (Mrk 335, PG 0844+349, PG 0953+414, PG 1211+143,
PG 1426+015), the photon index is too steep ($\Gamma>1.96$) to determine \mbh\
using the decay phase of GRO J1655-40.
\label{tab5}
\footnotesize
\end{table*}

%%%%%%%%%%%%%%%%%%% TAB6: CORRELATION ANALYSIS RESULTS %%%%%%%%%%%%%%%%
%\begin{sidewaystable}[h]
%\begin{landscape}
\begin{table*} 
%\rotate
\footnotesize
\caption{Statistical analysis results}
\begin{center}
\begin{tabular}{cccc} 
\hline        
\hline
\noalign{\smallskip}
Transition    &  Geometry   & $\langle M_{\rm BH,scal}/M_{\rm BH,RM}\rangle$  & $\langle max(M_{\rm BH,ratio})\rangle$\\
(1) & (2) & (3) & (4) \\
\noalign{\smallskip}
\hline 
\noalign{\smallskip}
GRO D05  & Sphere  & $1.5\pm0.4$  & $2.2\pm0.3$\\
\noalign{\smallskip}
               & Disk    &  $0.6\pm0.1$  & $3.3\pm0.5$\\
\noalign{\smallskip}
\hline
\noalign{\smallskip}
GRO R05  & Sphere  & $0.4\pm0.1$ & $5.7\pm1.4$\\
\noalign{\smallskip}
               & Disk    &  $0.16\pm0.03$ & $14.9\pm4.1$\\
\noalign{\smallskip}
\hline
\noalign{\smallskip}
GX D03   & Sphere  &  $2.2\pm0.5$ & $2.7\pm0.4$ \\
\noalign{\smallskip}
            & Disk    &  $0.8\pm0.2$ & $3.1\pm0.7$\\ 
\hline
\noalign{\smallskip}
GX R04   & Sphere  &  $0.7\pm0.1$ & $3.3\pm0.6$\\
\noalign{\smallskip}
            & Disk    & $0.3\pm0.1$ & $7.7\pm1.7$\\ 
\hline
\noalign{\smallskip}
XTE R98  & Sphere  &  $0.4\pm0.1$ &  $5.8\pm1.2$\\
\noalign{\smallskip}
 & Sphere-shift  & $1.1\pm0.2$ &  $2.5\pm0.4$\\
\noalign{\smallskip}
            & Disk    & $0.13\pm0.02$ &  $16.6\pm3.8$\\ 
\noalign{\smallskip}
\hline        
\hline
\end{tabular}
\end{center}
{\bf Columns Table 6}: 1= Reference source spectral transition. 2= 
Geometry of the soft photon emission region.
3= Ratio of the $M_{\rm BH}$ values determined
with the scaling method over the corresponding values from the reverberation mapping averaged
over the entire sample.  4= Maximum average ratio of the $M_{\rm BH}$ values which
provides an estimate of the average discrepancy (see Section 4.2 for details).
{\bf Notes}:  The ``sphere-shift" case corresponds to the spherical corona scenario
where the \mbh\ values obtained from the scaling technique have been multiplied
by a factor of 3. The large values of $\langle max(M_{\rm BH,ratio})\rangle$ for
patterns corresponding to rising phase outbursts (especially in the cases of
disk geometry) reflect the systematic underestimate of \mbh\ not the true departure
from the black hole mass ratio of 1.
\label{tab6}
\footnotesize
\end{table*}  
%\end{landscape}
%\end{sidewaystable}

%%%%%%%%%%%%%%%%%%% TAB7: CORRELATION ANALYSIS RESULTS %%%%%%%%%%%%%%%%
%\begin{sidewaystable}[h]
%\begin{landscape}
\begin{table*} 
%\rotate
\footnotesize
\caption{Correlation analysis results}
\begin{center}
\begin{tabular}{cccccc} 
\hline        
\hline
\noalign{\smallskip}
Transition    &  Geom.  & Slope   & Intercept & Spearman (Prob.) &  Kendall (Prob.) \\
(1) & (2) & (3) & (4) & (5) & (6) \\
\noalign{\smallskip}
\hline 
\noalign{\smallskip}
GRO D05  & Sphere  & $1.09\pm0.07$  & $-0.6\pm0.6$ &  0.78 ($9\times10^{-5}$)  & 0.59 ($4\times10^{-4}$) \\
\noalign{\smallskip}
               & Disk    & $1.03\pm0.06$  & $-0.6\pm0.5$ &  0.77 ($1\times10^{-4}$)  & 0.57 ($5\times10^{-4}$) \\
\noalign{\smallskip}
\hline
\noalign{\smallskip}
GRO R05  & Sphere  & $0.95\pm0.07$  & $-0.1\pm0.5$ &  0.78 ($5\times10^{-6}$)  & 0.62 ($2\times10^{-5}$) \\
\noalign{\smallskip}
               & Disk    & $0.89\pm0.07$  & $-0.1\pm0.5$ &  0.77 ($1\times10^{-5}$)  & 0.59 ($6\times10^{-5}$) \\
\noalign{\smallskip}
\hline
\noalign{\smallskip}
GX D03   & Sphere  & $0.93\pm0.08$  & $0.7\pm0.6$ &  0.79 ($4\times10^{-6}$)  & 0.63 ($2\times10^{-5}$)  \\
\noalign{\smallskip}
            & Disk    & $0.85\pm0.07$  & $0.9\pm0.6$ &  0.76 ($1\times10^{-5}$)  & 0.58 ($7\times10^{-5}$) \\ 
\hline
\noalign{\smallskip}
GX R04   & Sphere  & $1.03\pm0.07$  & $-0.5\pm0.5$ &  0.84 ($5\times10^{-7}$)  & 0.65 ($1\times10^{-5}$) \\
\noalign{\smallskip}
            & Disk    & $0.94\pm0.06$  & $-0.2\pm0.5$ &  0.82 ($2\times10^{-6}$)  & 0.64 ($2\times10^{-5}$) \\ 
\hline
\noalign{\smallskip}
XTE R98  & Sphere  & $0.97\pm0.09$  & $-0.4\pm0.7$ &  0.80 ($2\times10^{-6}$)  & 0.62 ($2\times10^{-5}$) \\
\noalign{\smallskip}
 & Sphere-shift & $0.99\pm0.08$  & $-0.03\pm0.6$ &  $\dots$ & $\dots$ \\
\noalign{\smallskip}
            & Disk    & $0.88\pm0.09$  & $-0.1\pm0.7$ &  0.80 ($2\times10^{-6}$)  & 0.62 ($2\times10^{-5}$) \\ 
\noalign{\smallskip}
\hline        
\hline
\end{tabular}
\end{center}
{\bf Columns Table 7}: 1= Reference source spectral transition. 2= 
Geometry of the soft photon emission region.
3= Best-fit
slope obtained fitting the data in Figures 3 and 4 with a straight line. 4= 
Best-fit intercept. 5= Spearman's $\rho$ rank correlation coefficient and 
related chance probability.
6= Kendalls's $\tau$ rank correlation coefficient and related chance probability.
{\bf Notes}: The quoted parameter errors for slope and intercept account for the
uncertainties on both x and y. The uncertainty on the average mass ratio represents
the spread around the mean $\sqrt{\sigma/n}$, where $n$ is the number of AGNs.
For the ``sphere-shift" case the values of the Spearman and Kendall analyses have been
omitted since they coincide with those of the ``sphere" case.
\label{tab6}
\footnotesize
\end{table*}  
%\end{landscape}
%\end{sidewaystable}

\end{document}